\documentclass[12pt]{iopart} 
\usepackage{amsfonts,amssymb}
\usepackage{iopams,cite}
\usepackage{graphicx}
\usepackage{color}
\usepackage{hyperref}

\newcommand{\text}[1]{\mbox{\scriptsize{#1}}}

\begin{document}
\title{Correlations and Symmetry of Interactions Influence Collective Dynamics of Molecular Motors}
\author{Daniel Celis-Garza, Hamid Teimouri and Anatoly B. Kolomeisky}
\address{Department of Chemistry and Center for Theoretical Biological Physics, Rice University, Houston, Texas, 77005, USA}
\eads{\mailto{tolya@rice.edu}}

\begin{abstract}

Enzymatic molecules that actively support many cellular processes, including transport, cell division and cell motility, are known as motor proteins or molecular motors. Experimental studies indicate that they  interact with each other and they frequently work together in large groups. To understand the mechanisms of collective behavior of motor proteins we study the effect of interactions in the transport of molecular motors along linear filaments. It is done by analyzing a recently introduced class of totally asymmetric exclusion processes that takes into account the intermolecular interactions via thermodynamically consistent approach. We develop a new theoretical method that allows us to compute analytically all dynamic properties of the system. Our analysis shows that correlations play important role in dynamics of interacting molecular motors. Surprisingly, we find that the correlations for repulsive interactions are  weaker and more short-range than the correlations for the attractive interactions. In addition, it is shown that symmetry of interactions affect dynamic properties of molecular motors.  The implications of these findings for motor proteins transport are discussed. Our theoretical predictions are tested by extensive Monte Carlo computer simulations.

\end{abstract}


\newpage

\section{Introduction}

Motor proteins or molecular motors are enzymatic molecules that actively participate in all major biological processes such as cellular transport, cell division, transfer of genetic information, synthesis of proteins, cell motility and  signaling \cite{howard_book,alberts_book,kolomeisky07,chowdhury13,kolomeisky13,kolomeisky_book}. They transform chemical energy from specific reactions that they catalyze (usually, hydrolysis or biopolymerization) into mechanical work  to support their functions. For example, the directed motion along linear cytoskeleton filaments by kinesin, myosin and dynein motor proteins is fueled by the hydrolysis of adenosine triphosphate (ATP) \cite{kolomeisky_book}. Biological molecular motors have been intensively studied in recent years, and currently the single-molecule dynamics of motor proteins is well described \cite{chowdhury13,kolomeisky13,veigel11}. Although the properties of individual molecules are very useful, in biological systems motor proteins typically function in large teams. This  underlines  the importance of understanding the collective behavior of molecular motors \cite{Uppulury12,kolomeisky13,Neri13}. 

Experimental studies of kinesin motor proteins moving along microtubules indicate that these molecular motors interact with each \cite{Roos08,Vilfan01,Seitz06}. It was argued that these interactions most probably are short-range and relatively weak attractive ($1.6\pm0.5 k_{B}T$) \cite{Roos08}. It is reasonable to assume that many other motor proteins have similar properties. At microscopic level, molecular motors are involved in a variety of chemical transitions such as binding to the filament, chemical transformations during the hydrolysis, dissociation from the track \cite{kolomeisky_book}. Intermolecular interactions influence all these processes, suggesting an important role for interactions in the collective behavior of molecular motors. However, the underlying mechanisms are still not well clarified \cite{kolomeisky13,Uppulury12}.  Existing theoretical studies of cooperative dynamics in interacting molecular motors are mostly phenomenological without quantitative description of relevant chemical processes \cite{Campas06,Pinkoviezky13,slanina08,Klumpp04}.

Recently, we introduced a new theoretical approach for analyzing collective properties of interacting molecular motors \cite{teimouri15}. It is based on using a class of non-equilibrium models called totally asymmetric simple exclusion processes (TASEP), which are very powerful for studying multi-particle dynamic phenomena \cite{Derrida98,Chou11,Bressloff13}. There are many processes in Chemistry, Physics and Biology that have been successfully analyzed by utilizing the asymmetric exclusion processes \cite{Chou11,Bressloff13,Frey03,Dong12,Golubeva12,tsekouras08,klumpp03}.  TASEPs were also employed before for investigating dynamic properties of motor proteins \cite{Chou11,Pinkoviezky13,Neri13,Dong12,Lipowsky01,Klumpp04}, including interacting molecular motors \cite{Pinkoviezky13,Klumpp04,Antal00,hager01}. But the main advantage of our method is a procedure that describes all chemical transitions at the single-molecule level using fundamental thermodynamic concepts \cite{teimouri15}. This allows us to properly couple microscopic properties of interacting molecular motors with  their collective dynamic behavior. 

Analyzing this theoretical approach, it was found that there is an optimal interaction strength (weakly repulsive) that can maximize the current through the system \cite{teimouri15}.  In addition, the calculations suggested that correlations play important role in dynamics of interacting molecular motors. However, the progress in understanding the cooperativity in motor proteins in this model was limited by the following issues. Two mean-field analytical treatments were proposed. But the first one, a simple mean-field approach (SMF), failed completely, as compared with extensive Monte Carlo computer simulations, producing unphysical trends in dynamic properties for strong interactions. The second approach, a cluster mean-field (CMF), worked better, but it involved very heavy numerical calculations. At the end, CMF was able to reproduce only qualitatively dynamics of interacting molecular motors and not even for all ranges of parameters. Furthermore, only symmetric splitting of interactions on hopping rates was considered. In addition, it was not possible to extend CMF to take into account more realistic features of motor protein's transport such as backward steps, bindings and unbindings from the filament, and more general symmetries of the interactions. To understand the mechanisms of cooperativity, it is important to have an analytical method that can successfully capture main features of interacting molecular motors, and which can be also  extended to more complex situations.   

In this paper,  a new theoretical framework for analyzing complex dynamics of interacting molecular motors via TASEP is presented. We develop a modified cluster mean-field (MCMF) approach that accounts for some correlations in the system. This provides a direct way of analytically calculating  all dynamic properties in the system, and the results agree quite well with computer simulations. The method allows us to explicitly analyze the role of interactions in dynamics of interacting molecular motors. More specifically, it is found that correlations are weaker and more short-range for repulsive interactions while for attractions they are stronger and more long-range. We also investigate the role of the symmetry of interactions and show that it might dramatically modify the dynamic behavior. But most importantly, the developed framework  allows us to understand the microscopic origin of dynamic phenomena in motor proteins and it can be easily generalized to account for more complex processes associated with molecular motors.

\section{Theoretical Description}\label{s:2}

\subsection{Model}

In our model, the transport of molecular motors along linear filaments is viewed as a motion of multiple particles on a lattice with $L$ ($L \gg 1$) sites, as shown in Fig. 1. The state of occupancy for each lattice site $i$ $(1 \le i \le L)$ is characterized by an occupation number $\tau_{i}$. If the site $ i $ is occupied then $ \tau_{i} = 1 $, if it is empty then $\tau_{i} = 0$. Each lattice site can accommodate only one particle.

In addition to exclusions, molecular motors can interact with each other via a short-range potential.  Here we assume that any two neighboring particles interact with each other with an energy $E$. The case of positive $E$ defines attractive interactions, while $E<0$ corresponds to repulsions. In other words, any bond connecting two neighboring particles on the lattice is associated with the energy $E$. In our system transition rates depend  if these bonds are broken or created. Any forward motion of the individual molecular motor that does not change the number of bonds is taking place with the rate 1. It can be done by a single molecule that do not have any neighbors (see Fig. 1), or it can involve breaking one bond and creating another one (Fig. 1).  In both cases, there is no energy change in the system.  However, the forward transition associated with creating a new bond has a rate $q \ne 1$. In this case, the molecular motor joins an existing cluster of particles: see Fig. 1. Similarly, the transition that is coupled with breaking the bond has a rate $r \ne 1$. Here the particle dissociates from the cluster but simultaneously it does not bind to another cluster ( Fig. 1). The transition rates $q$ and $r$ are associated with changes in energy. 

It has been argued that creating and breaking such bonds (or pairs of particles) can be viewed as opposing chemical transitions, which justifies the application of the detailed balance arguments \cite{teimouri15}. This leads to the following relation between the transitions rates,
\begin{equation}
\frac{q}{r}=\exp{\left(\frac{E}{k_{B}T}\right)}. 
\end{equation}
To evaluate  dynamic properties of molecular motors we need to know the explicit values for rates $q$ and $r$. The interaction energy can be split in the following way,
\begin{equation}\label{rates}
q=\exp{\left(\frac{\theta E}{k_{B}T}\right)}, \quad r = \exp{\left(\frac{(\theta-1)E}{k_{B}T}\right)},
\end{equation}  
where a dimensionless parameter $\theta$ ($0 \le \theta \le 1$) specifies how the energy affects these transition rates. Previously, only a symmetric splitting of interactions ($\theta=1/2$) has been considered \cite{teimouri15}. 

It is easy to understand the physical meaning of Eq. (\ref{rates}) \cite{teimouri15}. When the interactions between molecular motors are attractive ($E>0$), the rate of creating the bond is larger ($q>1$), while the rate of breaking the bond is smaller ($r<1$). For repulsive interactions ($E<0$) the trend is opposite --- it is faster to break the particle cluster ($r>1$) than to increase the cluster size ($q<1$). In the case of no interactions ($E=0$) these transitions rates are the same ($q=r=1$) and the model reduces to standard TASEP with only hard-core exclusions.

\begin{figure}[h]
\centering
\vspace{0.5cm}
\includegraphics[scale=0.5,height=3cm,width=15cm]{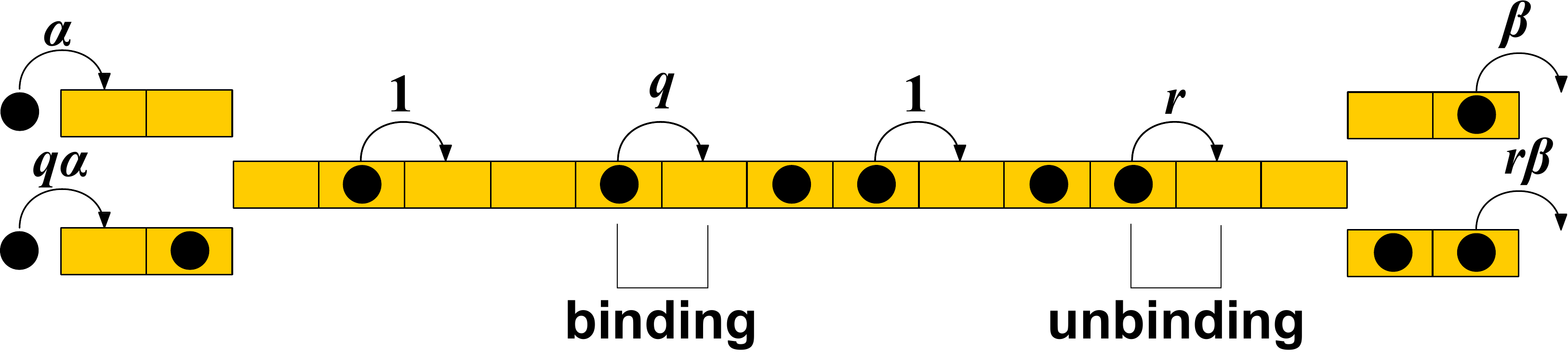}
\caption{Schematic view of the TASEP model for interacting molecular motors. Binding corresponds to particle joining the cluster, while unbinding describes the breaking from the cluster.}\label{fig1}
\end{figure}

In our model, particles enter the system from the left side of the lattice and they leave the system from the right side of the lattice: see Fig. 1. Interactions are also modifying the entrance and exit rates in comparison with the original TASEP model. When entering the system does not lead to creating a pair of particles the rate for this process is $\alpha$ (Fig. 1). However, entering with creating the bond has a rate $q \alpha$. Similarly, the exit rate for the case when no bond breaking is involved is equal to $\beta$, while the dissociating from the cluster is taking place with the rate $r \beta$ (Fig. 1).

\subsection{Modified Cluster Mean-Field Theory}

Previous theoretical studies indicated that correlations are important for the system with multiple particles \cite{teimouri15}. Neglecting correlations  leads to unphysical behavior for strong interactions between molecular motors \cite{teimouri15}. This indicates that any successful theoretical treatment must take correlations into account. This is the main idea of our approach that we call a {\it modified cluster mean-field}. 

To account for correlations we analyze bulk clusters of two neighboring sites on the lattice. Each cluster can be found in one of four possible states. We label a configuration with 2 empty sites as $(0,0)$, with two occupied sites as $(1,1)$, and two half-occupied clusters are labeled as  $(1,0)$ and $(0,1)$. Next, we introduce functions  $P_{11}$, $P_{10}$, $P_{01}$ and $P_{00}$ as probabilities of finding the configurations $(1,1)$, $(1,0)$, $(0,1)$ and $(0,0)$, respectively. The conservation of probability for these functions requires that 
\begin{equation}\label{normal}
P_{11}+P_{10}+P_{01}+P_{00}=1. 
\end{equation}
In addition, we have
\begin{equation}\label{P-rho}
P_{10}+P_{11}=\rho, \quad P_{01}+P_{11}=\rho,
\end{equation}
where $\rho$ is the bulk density (or the probability to find the particle at given site). Here it was assumed also that the bulk  density is uniform and independent of the position on the lattice if the two-site cluster is far away from the boundaries. Combining Eqs. (\ref{normal}) and (\ref{P-rho}) leads to $P_{10}=P_{01}$ and
\begin{equation}\label{normalization}
P_{10}+P_{00} + \rho = 1. 
\end{equation}

\begin{figure}[h]
\centering
\includegraphics[scale=0.5, height=6cm, width=6cm]{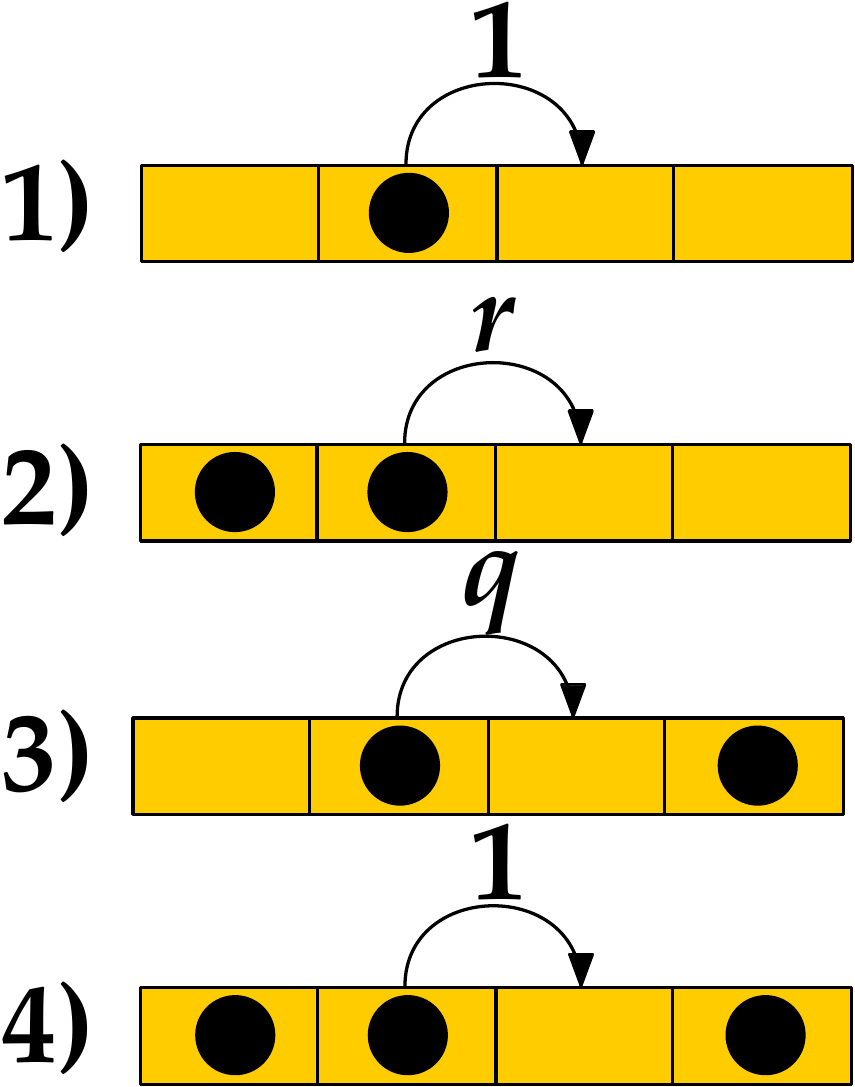}
\caption{Four-sites bulk lattice segments that are utilized for calculating the particle currents in the system.}\label{fig2}
\end{figure}

Let us consider a particle flux in the bulk of the system at large times when stationary conditions are achieved. We can concentrate on the segment of 4 consecutive sites as shown in Fig. 2. To measure the current, only transitions between the second and the third sites of the segment are counted. Then there are four possible configurations that support the particle current, as illustrated in Fig. 2. Correspondingly, the total flux  have 4 contributions from each configurations, $J_{bulk}=J_{1}+J_{2}+J_{3}+J_{4}$. The first contribution from configuration 1 (Fig. 2) can be written as
\begin{equation}\label{J1}
J_{1}=\gamma P_{10}\left(\frac{P_{00}}{\rho+ P_{00}}\right),
\end{equation} 
where $\gamma = \frac{1}{1+\exp{\left(\frac{E}{k_{B}T}\right)}}$. This expression is an approximation and it can be understood in the following way. The second factor ($P_{10}$) gives the probability that the cluster consisting of the second and third sites (see Fig. 2) is in the state $(1,0)$. The first factor ($\gamma$) gives a probability that the first site is empty, i.e., it is just a Boltzmann's factor. The third factor ($\frac{P_{00}}{\rho+ P_{00}}$) is a probability to have the last site empty. If we have the configuration $(1,0)$ in the middle cluster then the last site can be found in one of two states: it can be occupied with the probability $\rho$ or it can be empty with the probability $P_{00}$. This is because in this case the cluster consisting of the sites 3 and 4 can only be found in configurations $(0,0)$ or $(0,1)$. Then the particle current from the second configuration (Fig. 2) is equal to 
\begin{equation}\label{J2}
J_{2}=(1-\gamma) r P_{10}\left(\frac{P_{00}}{\rho+ P_{00}}\right).
\end{equation} 
Here $(1-\gamma)=\frac{\exp{\left(\frac{E}{k_{B}T}\right)}}{1+\exp{\left(\frac{E}{k_{B}T}\right)}}$ is a probability to have the first site occupied, and $r$ is the transition rate for this configuration. Similar arguments can be presented for contributions from configurations 3 and 4, yielding
\begin{equation}\label{J3}
J_{3}=\gamma q P_{10}\left(\frac{\rho}{\rho+ P_{00}}\right),
\end{equation} 
and 
\begin{equation}\label{J4}
J_{4}=(1-\gamma)  P_{10}\left(\frac{\rho}{\rho+ P_{00}}\right).
\end{equation} 

Combining together Eqs. (\ref{J1}), (\ref{J2}), (\ref{J3}) and (\ref{J4}) we obtain the expression for the total bulk current,
\begin{eqnarray}\label{current2}
J_{bulk} & = &\gamma\left(\frac{P_{10}P_{00}}{\rho+ P_{00}}\right) + (1-\gamma)r\left(\frac{P_{10}P_{00}}{\rho+ P_{00}}\right)  \nonumber \\ 
& & + q\gamma\left(\frac{\rho P_{10}}{\rho+ P_{00}}\right) + (1-\gamma)\left(\frac{\rho P_{10}}{\rho+ P_{00}}\right).
\end{eqnarray}
This equation can be also written in the following form,
\begin{equation} \label{current3}
J_{bulk} = A \left(\frac{P_{10}P_{00}}{1 - P_{10}}\right) + B \left(\frac{ \rho P_{10}}{1-P_{10}}\right),
\end{equation}
where auxiliary functions $A$ and $B$ are defined as
\begin{eqnarray}\label{A_B}
A = \frac{1 + r \exp{\left(\frac{E}{k_{B}T}\right)}}{1 + \exp{\left(\frac{E}{k_{B}T}\right)}},\quad B = \frac{q + \exp{\left(\frac{E}{k_{B}T}\right)}}{1 + \exp{\left(\frac{E}{k_{B}T}\right)}}.
\end{eqnarray}

To calculate explicitly dynamic properties in the system we have to express everything in terms of the bulk density $\rho$ and the interaction energy $E$. Eq. (\ref{normalization}) gives the connection between $P_{10}$, $P_{00}$ and $\rho$, and one more additional relation is needed in order to have all equations  only in terms of $\rho$ and $E$. We can approximate the function $P_{10}$ as
\begin{equation}\label{P10rho}
P_{10} \simeq \frac{\rho(1-\rho)}{1 - \rho + \rho \exp{\left(\frac{E}{k_{B}T}\right)}}.
\end{equation}
The physical meaning of this approximation can be explained if we note that $P_{10}$ is the probability to have the two-site cluster in the configuration $(1,0)$. This probability is equal to the product of two terms: one is the probability to have the first site occupied ($\rho$) and the second term is the probability that the second site is empty {\it given} that the first one is not ($(1-\rho)/[1 - \rho + \rho \exp{\left(\frac{E}{k_{B}T}\right)}]$). It can be argued that the situation when two sites in the cluster are occupied is affected by the interactions between them, and we approximate it via the usual Boltzmann's factor. One can see that this equation leads to a very reasonable behavior at the limiting cases. When there is no interactions ($E=0$) it predicts that $P_{10}=\rho(1-\rho)$, as expected. For very strong repulsions ($E \rightarrow -\infty$) it gives $P_{10}=\rho$, which is a correct result since in this limit our problem is identical to the motion of non-interacting dimers on the lattice \cite{teimouri15,Lakatos03}. For very large attractions ($E \rightarrow \infty$) the prediction is that $P_{10} \rightarrow 0$. This is again seems to be a reasonable result because in this case one is expecting to have the whole system fully occupied without any vacancies.

Finally, taking into account all approximations we obtain the general expression for the bulk current only in terms of the particle density $\rho$ and the interaction $E$,
\begin{eqnarray}\label{current_ro}
J_{bulk} & = & \frac{A\rho(1-\rho)^2\left[1-2\rho+\rho \exp{\left(\frac{E}{k_{B}T}\right)}\right]}
{\left[(1-\rho)^{2}+\rho \exp{\left(\frac{E}{k_{B}T}\right)}\right]\left[1-\rho+\rho \exp{\left(\frac{E}{k_{B}T}\right)}\right]}+ \nonumber \\ & &   \frac{B\rho^{2}(1-\rho)}{\left[(1-\rho)^{2}+\rho \exp{\left(\frac{E}{k_{B}T}\right)}\right]} 
\end{eqnarray}
For the case of zero interactions this equation suggests that $J_{bulk}=\rho(1-\rho)$, the known result form the original TASEP model \cite{Derrida98,Chou11}. For strong repulsions ($E \rightarrow -\infty$) we predict that 
\begin{equation}\label{current_strong_repul}
J_{bulk} = \frac{\rho(1 - 2\rho)}{1 - \rho}.
\end{equation}
This is identical to the expression that was derived earlier for TASEP of dimers \cite{Lakatos03}. For large attractions ($E \rightarrow \infty$) it predicts that the bulk current vanishes, $J_{bulk} = 0$, and this is expected because  particles will not be able to move since they will be stuck together in one large cluster.

At boundaries the dynamics in the system is governed by exit and entrance rates. Using the same approximations as explained above for the bulk fluxes it can be shown that the expressions for entrance current is given by
\begin{equation}\label{current_ld2}
J_{entr} = \frac{\alpha(1-\rho)\left[1-2\rho+\rho\exp{\left(\frac{E}{k_{B}T}\right)}\right]+\alpha q\rho(1-\rho)}{1 - \rho + \rho \exp{\left(\frac{E}{k_{B}T}\right)}}.
\end{equation}
For $E=0$ this equation reduces to  $J_{entr}=\alpha(1-\rho)$, which is expected for this situation. For $E \rightarrow - \infty$it predicts that $J_{entr}=\alpha(1-2\rho)$, in agreement with known results on TASEP of extended objects \cite{Lakatos03}. For strong attractions ($E \rightarrow \infty$) the current disappears, $J_{entr} \rightarrow 0$.  Similarly, for the exit current we obtain
\begin{equation}\label{current_rb2}
J_{exit} = \frac{\beta \rho\left [1-\rho + r\rho \exp{\left(\frac{E}{k_{B}T}\right)}\right]}{1 - \rho + \rho \exp{\left(\frac{E}{k_{B}T}\right)}}.
\end{equation}
Again, for $E=0$ and for $E \rightarrow - \infty$ it produces the expected result, $J_{exit}= \beta \rho$, while for strong attractions it leads  to $J_{exit} \rightarrow 0$.

\subsection{Phase Diagrams}\label{ss:phase}

Similarly to the original TASEP, it can be argued that in the system of interacting molecular motors there are three dynamic phases at stationary conditions. When the rate limiting step is the entrance into the system we have a low-density (LD) phase. For the case of exiting being small a high-density (HD) phase will be realized. Finally, when bulk processes are the most important, the system is in a maximal-current (MC) phase.     

Our analytical theory can calculate explicitly the phase boundaries. The MC phase is characterized by a condition that $ \frac{\partial J_{bulk}}{\partial \rho} = 0 \label{rho_max}$, which leads to the following expression:
\begin{eqnarray}\label{mc_ploy1}
\hspace{-2.5cm}\left(2 \rho ^2-4 \rho +1\right) (\rho -1)^4-\rho ^4 \exp\left((\theta +3) \frac{E}{k_{B}T} \right)-\rho ^2 (2 \rho -1) (\rho -1)^2 \exp\left((\theta +2) \frac{E}{k_{B}T} \right)\nonumber\\
\hspace{-2.5cm}+(\rho -1)^6 \exp\left(\theta \frac{E}{k_{B}T} \right)-(\rho -2) \rho  (\rho -1)^4 \exp\left((\theta+1) \frac{E}{k_{B}T} \right)+\left(\rho ^4-2 \rho ^5\right) \exp\left(4 \frac{E}{k_{B}T} \right)\nonumber\\
\hspace{-2.5cm}-\rho  \left(4 \rho ^3-16 \rho ^2+15 \rho -4\right) (\rho -1)^2 \exp\left(\frac{E}{k_{B}T} \right) -\rho ^3 \left(\rho ^3-10 \rho ^2+13 \rho -4\right) \exp\left(3 \frac{E}{k_{B}T} \right)\nonumber\\
\hspace{-2.5cm}+\rho ^2 \left(3 \rho ^4-22 \rho ^3+39 \rho ^2-26 \rho +6\right) \exp\left(2 \frac{E}{k_{B}T} \right) = 0.
\end{eqnarray} 
For $E=0$  this complicated expression reduces to the following  formula,
\begin{equation}
-4 \rho ^5+10 \rho ^4-16 \rho ^3+14 \rho ^2-8 \rho +2 = 0,
\end{equation}
which has only one real root, $ \rho = \frac{1}{2} $. For very large repulsions ($E \rightarrow -\infty$) one can obtain from Eq.(\ref{mc_ploy1}),
\begin{equation}
\left(2 \rho ^2-4 \rho +1 \right) (\rho -1)^4=0.
\end{equation}
This equation has three roots but only one of them is physically reasonable, $\rho=1-1/\sqrt{2}$, which leads to a nonzero flux in the system. (The root $\rho=1$ does not support the nonzero flux through the system and it can be neglected.) Substituting this density  into Eq. (\ref{current_strong_repul}) leads to a prediction that the particle flux in this case is equal to $J=3-2 \sqrt{2} \approx 0.17$. For large attractions ($E \rightarrow \infty$)  Eq.(\ref{mc_ploy1}) predicts that $\rho=1/2$, but the current here is approaching zero, as was discussed above. For general conditions, this equation can be always solved  numerically and after choosing the physically relevant root for the density in the MC phase, $\rho_{MC}$, the particle fluxes can be calculated using Eq. (\ref{current_ro}). The phase diagrams calculated via this method are presented in Fig. 3 for various sets of parameters.

\begin{figure}[h]
\centering
\includegraphics[clip,width=0.65\textwidth]{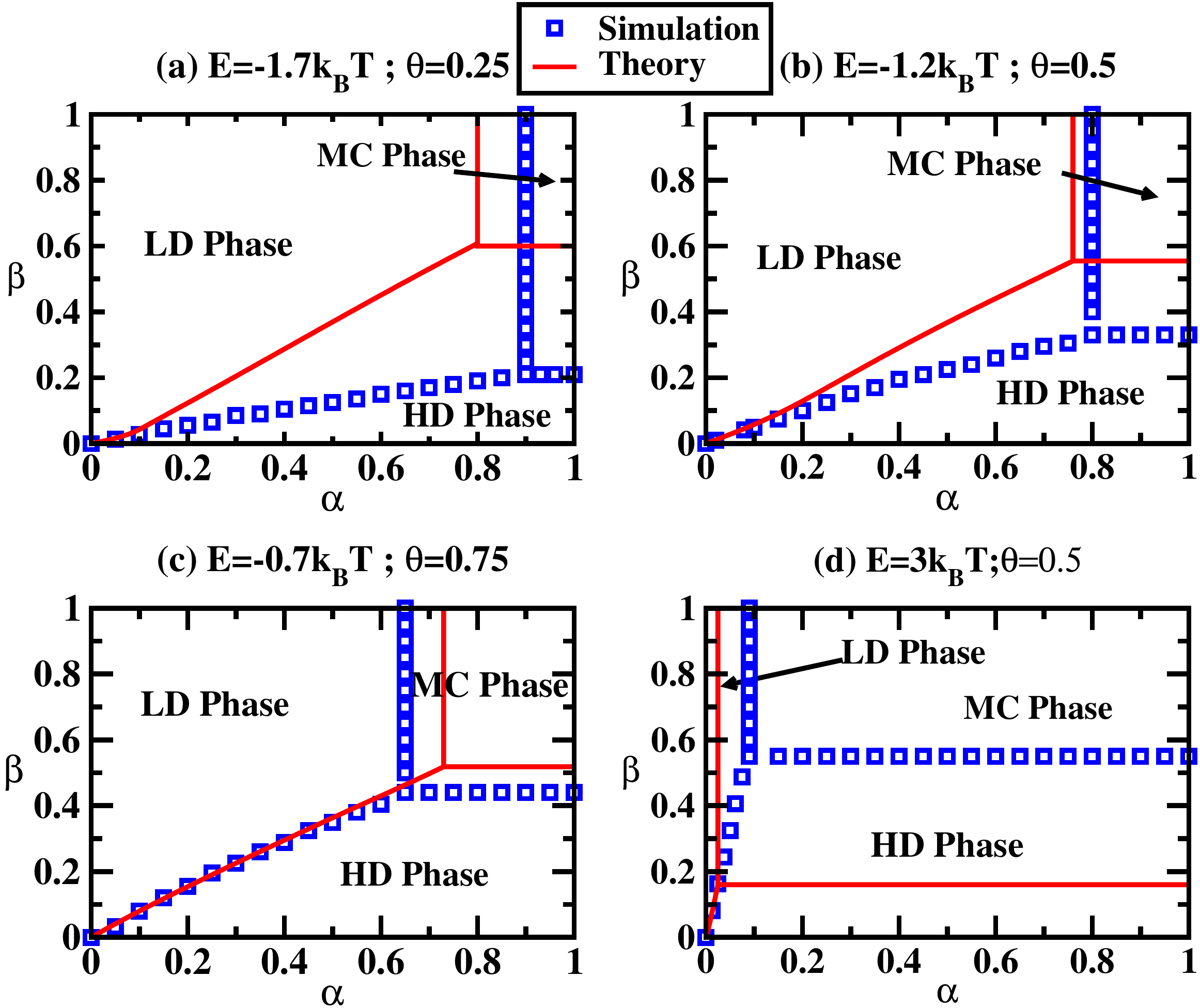}
\caption{Stationary phase diagrams of TASEP with interacting particles for various interaction strengths and interaction splittings:  (a) $E = -1.7$k$_{B}$T, $\theta=0.25$; (b) $E = -1.2$k$_{B}$T, $\theta=0.5$; (c) $E = -0.7$k$_{B}$T, $\theta=0.75$; (d) $E = 3$k$_{B}$T, $\theta=0.5$. Lines are theoretical predictions, symbols are from Monte Carlo computer simulations. }
\end{figure}

To determine the  density of molecular motors in the LD phase and the boundary lines separating the low-density and the maximal-current phases, we use the continuity of the stationary currents at the transition line, $J_{bulk}=J_{entr}$. Combining Eqs. (\ref{current_ro}) and (\ref{current_ld2}) yields the following expression for  the entrance rate $\alpha$,
\begin{equation}\label{e:ac}
 \hspace{-2cm} \alpha  =  \frac{A\rho(1-\rho)\left[1-2\rho+\rho \exp{\left(\frac{E}{k_{B}T}\right)}\right]+B\rho^{2}\left[1 - \rho + \rho \exp{\left(\frac{E}{k_{B}T}\right)}\right]}{\left[1-2\rho+\rho\exp{\left(\frac{E}{k_{B}T}\right)}+ q\rho\right]\left[(1-\rho)^{2}+\rho \exp{\left(\frac{E}{k_{B}T}\right)}\right]}.
\end{equation}
Solving this equation for $\rho$ provides an estimate for the particle density in the bulk of the system in the LD phase. Increasing the entrance rate $\alpha$ leads to large bulk densities, and the phase boundary between LD and MC phase is achieved when $\rho=\rho_{MC}$. For example, for zero interactions, $E=0$, from Eq. (\ref{e:ac}) it follows that $\rho_{LD}=\alpha$, and the phase boundary corresponds to $\alpha=0.5$. These estimates fully agree with results from the original TASEP with non-interacting particles \cite{Derrida98,Chou11}. For the case of very strong repulsions, $E \rightarrow -\infty$, we derive from Eq. (\ref{e:ac}) that $\rho_{LD}=\alpha/(1+\alpha)$, and the phase boundary between LD and MC phase corresponds to $\alpha=\sqrt{2}-1 \approx 0.41$. This again agrees with known results on TASEP of extended objects \cite{Lakatos03}. In the opposite limit of very large attractions, $E \rightarrow \infty$, the low-density phase cannot be realized for any nonzero values of $\alpha$.

Similar calculations can be performed for obtaining properties of the HD phase and the boundaries between high-density and maximal-current phase. The exit rate $\beta$ is coupled with the  density $\rho$ via
\begin{equation}\label{e:bc}
\hspace{-2cm} \beta =\frac{A (1-\rho)^{2}\left[1- 2\rho + \rho \exp{\left(\frac{E}{k_{B}T}\right)}\right] + B \rho (1 - \rho)\left[1- \rho + \rho \exp{\left(\frac{E}{k_{B}T}\right)}\right]}
{\left[(1-\rho)^{2} + \rho \exp{\left(\frac{E}{k_{B}T}\right)}\right] \left[1-\rho + r \rho \exp{\left(\frac{E}{k_{B}T}\right)}\right]}.
\end{equation}
This expressions allows us to calculate the bulk particle density in the HD phase. One can see that increasing the exit rate $\beta$ lowers the bulk density until the phase boundary with the MC phase is reached at $\rho=\rho_{MC}$. This can be illustrated by again considering the limiting cases. When motor proteins do not interact with each other ($E=0$) Eq.(\ref{e:bc}) yields $\rho_{HD}=1-\beta$ and the phase boundary between HD and MC phase can be found at $\beta=0.5$. This is fully consistent with known results for TASEP of non-interacting particles \cite{Derrida98,Chou11}. For strong repulsions we predict that $\rho_{HD}=(1-\beta)/(2-\beta)$, and the phase boundary between HD and MC phases is observed at $\beta=2-\sqrt{2} \approx 0.59$. Note that these results slightly differ from calculations for TASEP of dimers because of the different exiting rules \cite{Lakatos03}. For strong attractions the flux through the system is vanishing and the high-density is always observed.

The phase boundary between LD and HD phases can be estimated from the condition that at this line the particle currents from both phases become equal, $J_{LD} = J_{HD}$. It can be shown from Eqs. (\ref{current_ld2}) and (\ref{current_rb2}) that
\begin{equation}\label{beta_ld-hd}
\hspace{-3cm} \frac{\beta}{\alpha}=\left[\frac{q\rho_{LD}(1-\rho_{LD})+(1-\rho_{LD})(1-2\rho_{LD}+\rho_{LD}\exp{\left(\frac{E}{k_{B}T}\right)})}{\rho_{HD}(1-\rho_{HD} + r\rho_{HD}\exp{\left(\frac{E}{k_{B}T}\right)})}\right]\left[\frac{1 - \rho_{HD} + \rho_{HD} \exp{\left(\frac{E}{k_{B}T}\right)}}{1 - \rho_{LD} + \rho_{LD} \exp{\left(\frac{E}{k_{B}T}\right)}} \right].
\end{equation}
In this expression, densities $ \rho_{HD} $ and $ \rho_{LD} $ are obtained by solving Eqs. (\ref{e:ac}) and (\ref{e:bc}), respectively, for specific values of $E$ and $ \theta $. For $E = 0$ we find that the LD-HD phase boundry is given by $\beta = \alpha $, and the triple point (where LD, HD and MC phases meet together) is found at $ \beta_{c} = \alpha_{c} = 0.5$, as expected for the standard TASEP  \cite{Derrida98,Chou11}.

\begin{figure}[h]
\centering
\includegraphics[clip,width=0.65\textwidth]{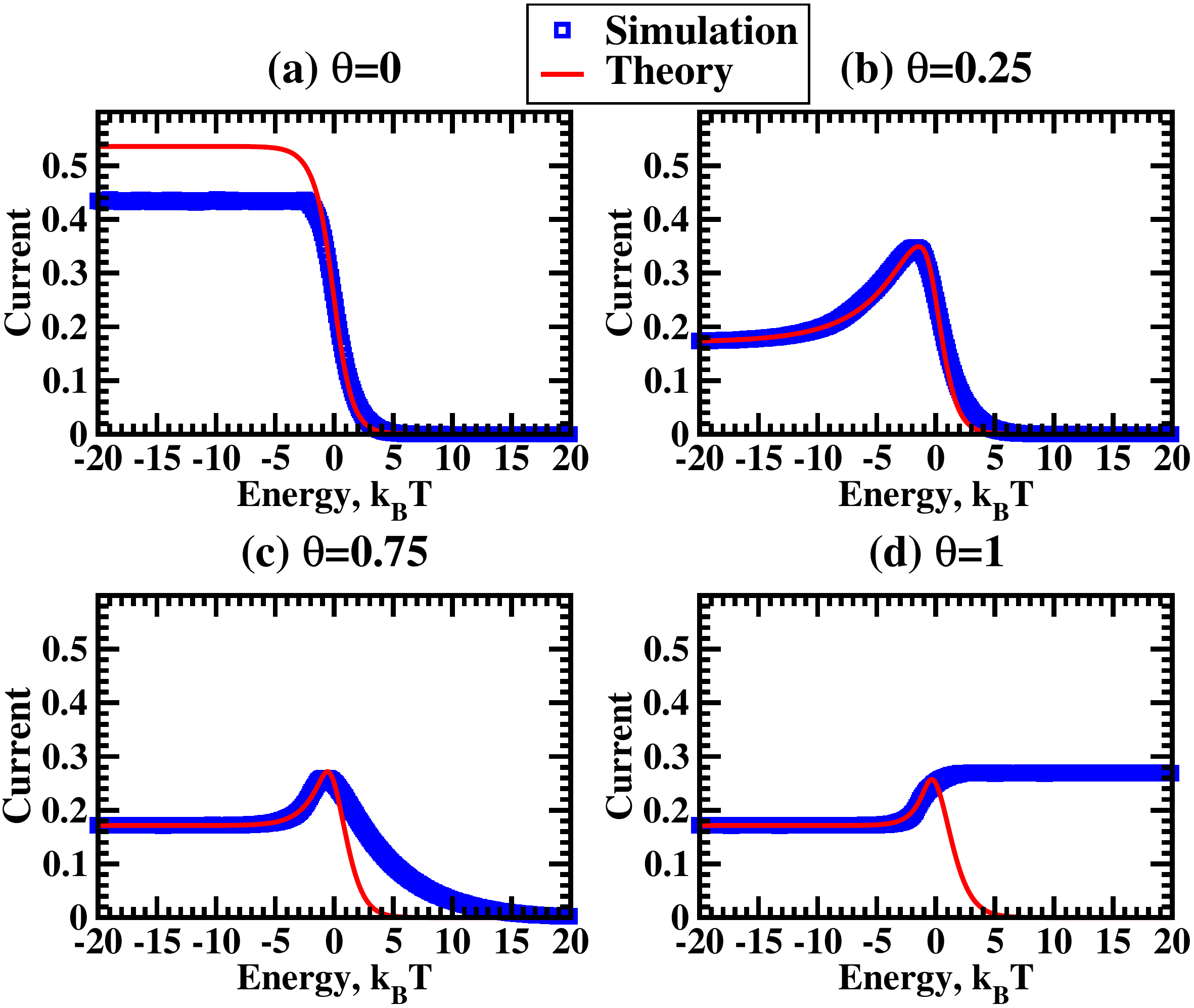}
\caption{Maximal particle currents as a function of the interaction energy for different energy splittings: (a) $\theta=0$, (b) $\theta=0.25$, (c) $\theta=0.75$ and (d) $\theta=1$. In simulations  $\alpha=\beta=1$ was utilized.}
\end{figure}

\section{Monte Carlo Simulations and Discussions}\label{s:3}

Because of the approximate nature of our method, it is important to test these  theoretical predictions. It was done in this work by running extensive computer Monte Carlo simulations.  We utilized the Monte Carlo algorithm known as a random sequential update. In our simulations we used a lattice of size $L=1000$ to minimize any finite size and boundary effects. The particle current and density profiles of molecular motors  were averaged over $10^{8}$ Monte Carlo steps. To ensure that the system is at the stationary-state conditions, the first $20\%$ of events were discarded. We have used a precision of $0.01$ when comparing density profiles to construct accurate phase diagrams. The error in calculating the phase boundaries by our method was estimated to be less than $1\%$.

In Fig. 3 we compare theoretically calculated phase diagrams with results obtained in Monte Carlo computer simulations. One can see that for relatively weak interactions theory agrees quite well with computer simulations (Fig. 3c), while for stronger interactions (attractive or repulsive) the agreement is mostly qualitative (although still better for repulsions): see Figs. 3a, 3b and 3d. Comparing  phase behavior at different interactions, one can notice that the LD phase is dominating at repulsions, but the HD phase is more prevalent for attractions. These observations are consistent with expectations that repulsions lead to smaller particle clusters and lower density while attractions stimulate the formation of large clusters, which corresponds to higher density. 

However, our method works much better for predicting particle fluxes in the MC phase, as illustrated in Fig. 4. The theory correctly describes the fluxes for repulsive interactions for all ranges of parameters (with the exception of the special case corresponding to $\theta=0$). But for attractions the good agreement is found only for small $\theta$. For larger values of the energy splittings ($\theta > 0.25$) there is only a qualitative agreement on the overall trends: the fluxes decrease to zero with increasing the interaction strength (with the exception of the special case corresponding to $\theta=1$). 

These observations suggest that correlations are important for understanding dynamic properties of interacting molecular motors. To quantify this effect, we investigated  a correlation function $C_{i}$ defined as
\begin{equation}\label{sim_corr}
C = \langle \tau_{i} \tau_{i+1} \rangle - \langle \tau_{i} \rangle \langle \tau_{i+1} \rangle ,\quad i=1,...,L-1
\end{equation}
where two-point and one-point density functions are given by
\begin{eqnarray}\label{two_point}
\langle \tau_{i} \tau_{i+1} \rangle=\sum\limits_{\tau_{i}} \sum\limits_{\tau_{i+1}}\tau_{i}\tau_{i+1} P(\tau_{i},\tau_{i+1})=P_{11},\\
\langle \tau_{i} \rangle=\sum\limits_{\tau_{i}} \tau_{i}P(\tau_{i})=\rho.
\end{eqnarray}
The physical meaning of the correlation function $C_{i}$ is that it gives a measure of how the presence of the particle at site $i$ affects the occupation of the neighboring site $i+1$. Using the definition together with the normalization condition and Eq. (\ref{P10rho}), we obtain the following analytical expression for $C$, 
\begin{equation}\label{correlation}
C(E)=\frac{\rho^{2}(1- \rho)\left[\exp \left(\frac{E}{k_{B} T} \right)-1\right]}{1+\rho \left[\exp \left(\frac{E}{k_{B} T} \right)-1 \right]}.
\end{equation}
Note that $C$ is uniform in the bulk of the system. For the case of zero interactions ($E=0$) it predicts that $C=0$. This fully agrees  with what we know about TASEP for noninteracting particles. Here a simple mean-field theory, that completely neglects any correlations, works quantitatively  well and it correctly describes the majority of dynamic properties of the system \cite{Derrida98,Chou11}. For strong repulsions ($E \rightarrow -\infty$) it gives  $C=-\rho^{2}$, while for strong attractions ($E \rightarrow \infty$) we have $C=\rho(1-\rho)$.

\begin{figure}[h]
\centering
\includegraphics[clip,width=0.67\textwidth]{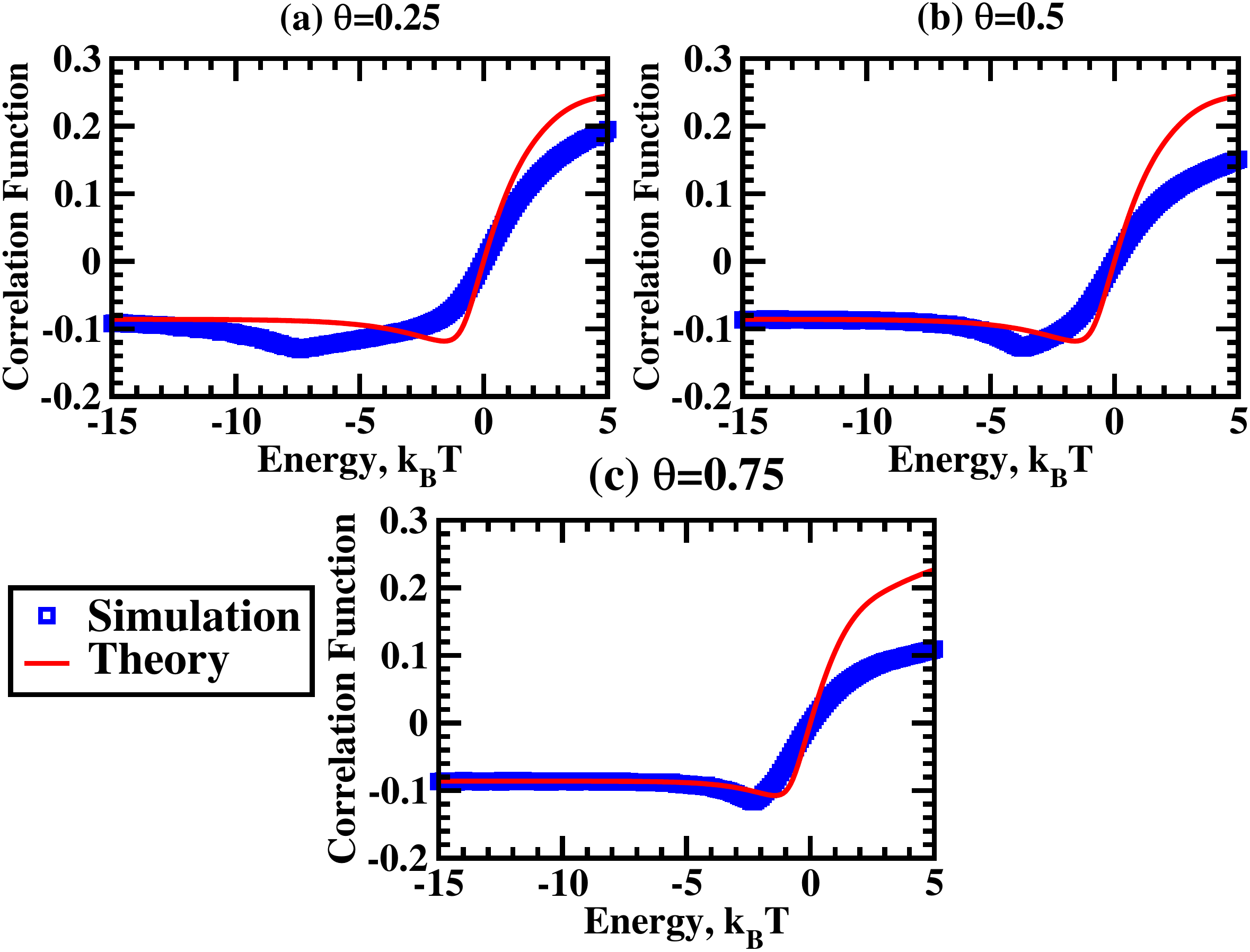}
\caption{Correlations as a function of the interaction energy for: (a) $\theta=0.25$, (b) $\theta=0.5$, (c) $\theta=0.75$. In simulations  $\alpha=\beta=1$ was utilized.}
\end{figure}

The correlation functions predicted in our method and obtained from computer simulations are presented in Fig. 5. It is interesting to analyze these data. The physical meaning of the correlation function $C_{i}$ is that it gives a measure of how the presence of the particle at site $i$ affects the occupation of the neighboring site $i+1$. When there are no correlations we have $C=0$. Negative correlation functions ($C <0$)  indicate that there is a less probability to find the particle next to the already occupied site. This is the case for repulsive interactions. In contrary, positive values for $C$ suggest that the presence of the particle at given site enhances the probability to find the particle at the neighboring site. It is clear that this situation can be realized for attractive interactions. Comparing theoretical predictions with Monte Carlo simulations (Fig. 5) again indicates that our theory works very well for repulsive interactions, while for attractions, although the trends are correctly picked up, there are deviations. 

The analysis of results presented in Figs. 3, 4 and 5 strongly indicates that correlations are important for understanding the mechanisms of interacting molecular motors. However, it also raises a question of why our theoretical approach, that explicitly takes into account some correlations, is able to correctly describe the stationary properties only for repulsive interactions and for weak attractions. To answer this question we note that the dynamic behavior strongly depends on the sign of the interactions. For $E<0$, the presence of the particle at the site $i$ leads to a lower probability of finding another particle at the site $i+1$. Then if there is nothing at the site $i+1$ the occupancy of the site $i+2$ will be independent of the fact that there is the particle at the site $i$.  These arguments suggest that correlations for repulsive interactions are {\it short-range} and relatively weak. For $E>0$ the situation is very different. Here the presence of the particle at the site $i$ stimulates the occupancy of the site $i+1$, and consequently the occupation state of the site $i+2$ depends on the state of the site $i$. This is consistent with {\it long-range} and strong correlations. By construction (see Sec. 2.2), our theory accounts only for short-range correlations because the evolution of two-site clusters is monitored. This is the main reason why our approach is so successful for repulsive interactions, while providing mostly a qualitative description for attractive interactions.  

One of the main advantages of our theoretical method is the fact that it can be easily extended to account for more realistic features of the motor proteins transport. To illustrate this we analyze the effect of varying the splitting coefficient $\theta$ on multi-particle dynamics of interacting molecular motors. The results are presented in Figs. 4 and 6. One can see that dynamics is different for small and large values of the interaction splittings.  It can be concluded from Eq. (2) that small $\theta$ describe the situation when the formation of the particle clusters is weakly affected by the interactions. At the same time, the breaking of the particle bonds is strongly influenced by interactions. For $\theta \approx 1$ the trend is opposite: the particle cluster formation depends strongly on interactions, while the bond breaking is not. 

Fig. 4 shows that the particle current (in the MC phase) is generally a non-monotonic function of interactions. At large repulsions  the current saturates, while for large attractions the fluxes are going to zero. The maximal particle current is achieved for relatively weak repulsive interactions ($E \simeq -(1-2)$ k$_{B}$T). The monotonic decrease in the particle current is only observed for the special case of $\theta=0$. Similar dynamics is observed for large $\theta >0.9$ (see Fig. 6), but here the most optimal conditions are reached  now for positive interactions. The case of $\theta=1$ is again a special one, and the monotonic increase in the current is observed for all range of interactions.

\begin{figure}[h]
\centering
\includegraphics[clip,width=0.67\textwidth]{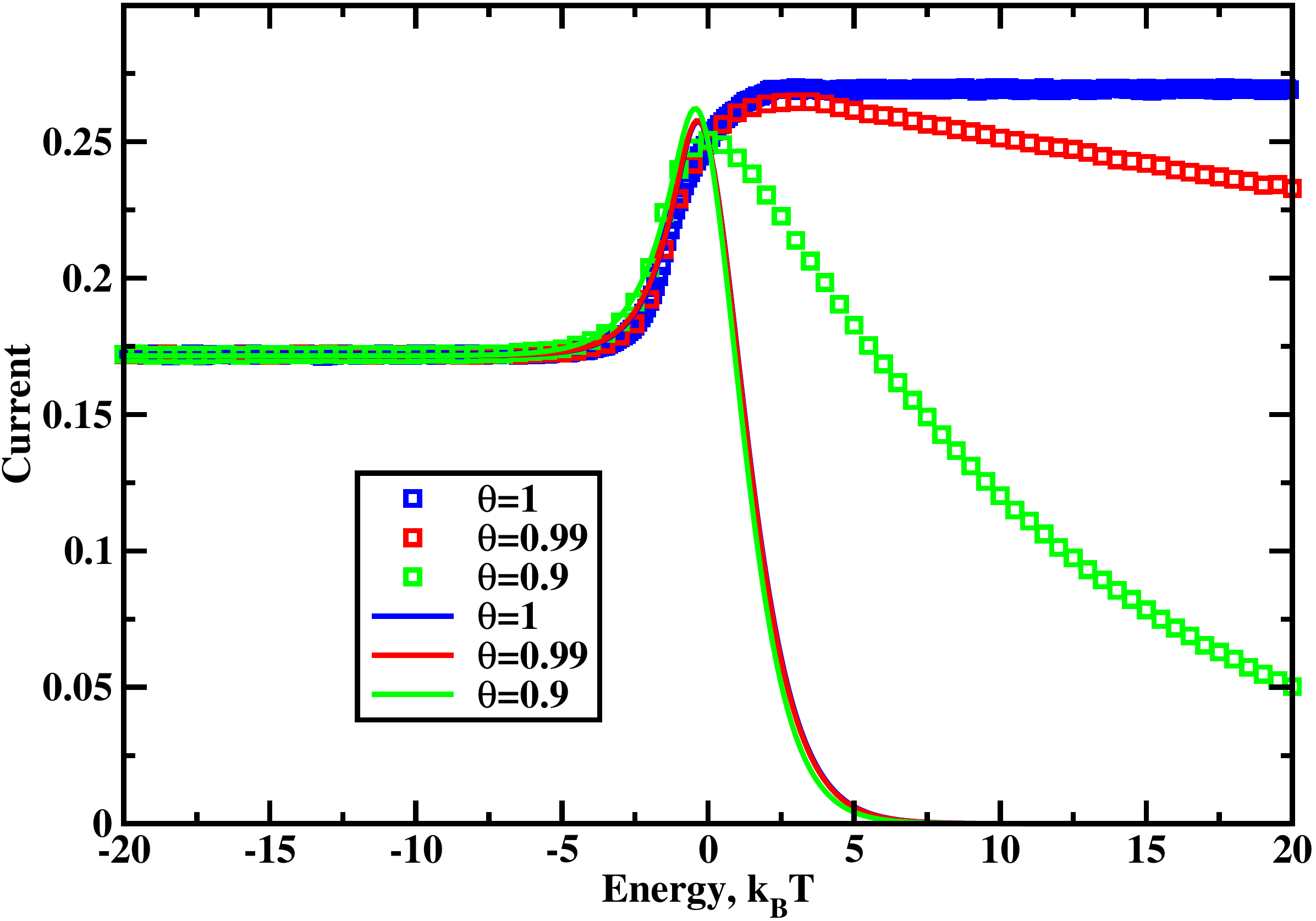}
\caption{Maximal particle currents as a function of the interaction energy for large energy splittings. Lines are theoretical predictions, symbols are from computer simulations.}
\end{figure}

In light of these findings, it is important to discuss the effect of intermolecular interactions for real motor proteins. These interactions have been measured experimentally for kinesins, indicating weak attractions of order of $E=(1.6 \pm 0.5)$ k$_{B}$T \cite{Roos08}.  Previous theoretical studies suggested that kinesins function at the conditions that do support the maximal current, but the analysis was based on the symmetric splitting of interactions for transitions rates ($\theta=0.5$) \cite{teimouri15}. Our new results presented in Figs. 4 and 6 indicate that this is probably a reasonable description of multi-particle dynamics of kinesins for for most interaction splittings ($\theta <0.9$). In this case the kinesin might operate at the conditions where small changes in interactions lead to large modification in the particle dynamics. It has been argued that this might be important for maintaining robust cellular transport \cite{kolomeisky13,Uppulury12,teimouri15}. However, our results (Fig. 6) also suggest another intriguing possibility that the kinesin fluxes might be optimized if the splitting affects more the formation of particle clusters ($\theta>0.9$). The parameter $\theta$ is a microscopic property that cannot be obtained from our mesoscopic theoretical method. To test this idea it will be important to measure and calculate this quantity in more advanced experimental and theoretical investigations.

\section{Summary and Conclusions}\label{s:4}

We developed a new theoretical approach to analyze the role of intermolecular interactions in the collective dynamics of molecular motors that move along linear filaments. Our method is based on utilizing totally asymmetric exclusion processes, which have been successfully applied for studying various processes in Chemistry, Physics and Biology.   It modifies the  transition rates by interactions via fundamental thermodynamic arguments.  A simple theoretical framework, that we call the modified cluster mean-field and that takes into account some correlations, is presented and fully discussed. It allows us to calculate analytically or numerically exactly all dynamic properties of interacting molecular motors. We find that interactions induce correlations in the system of collectively moving motor proteins, and the strength of correlations depends on the sign of the interactions. It was argued that for repulsions the correlations are short-range and relatively weak, while for attractions the range and amplitude of correlations are larger. This also leads to different dynamic behavior of interacting molecular motors. For repulsions the dynamics is weakly affected by the strength of interactions, however for attractions the dynamics is modified much stronger. We also investigated the effect of the symmetry of interaction by analyzing splittings between different transitions. It was found that when the formation of particle clusters is weakly affected by interactions the most optimal fluxes can be realized for weakly repulsive interaction. But when breaking of bonds between neighboring particles is strongly influenced, the maximal current can be achieved for  attractive interactions. Furthermore, the importance of these results for kinesin motor proteins has been discussed.

The most important advantage of our method is that it can be easily extended to investigate additional more realistic features of molecular motors transport such as backward steps, binding and unbinding of motor proteins at all sites, multiple parallel pathways and limited resources of motor proteins in the surrounding solution. It will be very interesting to generalize our approach to study these phenomena if we want to understand better mechanisms of cellular transport processes. However, despite its simplicity and the successful application for the interacting molecular motors, it should be noted that our method is still approximate and many important features are not well described. For example, for  a large range of parameters the effect of attractive interactions is given only qualitatively. Thus it will be important to test our theory in experiments and in more advanced theoretical models.

\section*{Acknowledgments}

The work was supported by grants from the Welch Foundation (C-1559), from the NSF (Grant CHE-1360979), and by the
Center for Theoretical Biological Physics sponsored by the NSF (Grant PHY-142765).

\end{document}